\newcommand {\ignore}[1]{}

\newcommand{\noi}{\noindent}
\newcommand{\bc}{\begin{center}}
\newcommand{\ec}{\end{center}}

\newlength\lah
\newlength\sah

\def\ifmath#1{\relax\ifmmode #1\else $#1$\fi}
\def\3quarter{{\textstyle{3 \over 4}}}

\def\ra{\rightarrow}

\def\lf{\leaders\hbox to 1em{\hss.\hss}\hfill}

\def\21{$SU(2) \ot U(1)$}

\def\O{\hbox{$\cal O$ }}

\def\neu{\hbox{neutrino }}

\def\eq#1{{eq. (\ref{#1})}}

\def\lsim{\raise0.3ex\hbox{$\;<$\kern-0.75em\raise-1.1ex\hbox{$\sim\;$}}}
\def\gsim{\raise0.3ex\hbox{$\;>$\kern-0.75em\raise-1.1ex\hbox{$\sim\;$}}}

\def\bel{\begin{letter}}
\def\eel{\end{letter}}
\def\beq{\begin{equation}}
\def\eeq{\end{equation}}
\def\bef{\begin{figure}}
\def\eef{\end{figure}}
\def\bet{\begin{table}}
\def\eet{\end{table}}
\def\bea{\begin{eqnarray}}
\def\ba{\begin{array}}
\def\ea{\end{array}}
\def\bi{\begin{itemize}}
\def\ei{\end{itemize}}
\def\ben{\begin{enumerate}}
\def\een{\end{enumerate}}
\def\ra{\rightarrow}
\def\ot{\otimes}

\def\eea{\end{eqnarray}}
\def\arnps#1#2#3{        {\it Ann. Rev. Nucl. Part. Sci. }{\bf #1} (19#2) #3}

\def\np#1#2#3{           {\it Nucl. Phys. }{\bf #1} (19#2) #3}
\def\pl#1#2#3{           {\it Phys. Lett. }{\bf #1} (19#2) #3}
\def\pr#1#2#3{           {\it Phys. Rev. }{\bf #1} (19#2) #3}
\def\prep#1#2#3{         {\it Phys. Rep. }{\bf #1} (19#2) #3}
\def\prl#1#2#3{          {\it Phys. Rev. Lett. }{\bf #1} (19#2) #3}

\def\zp#1#2#3{           {\it Zeit. fur Physik }{\bf #1} (19#2) #3}

\def\n.c.#1#2#3{         {\it Nuovo Cim. }{\bf #1} (19#2) #3}
\def\r.n.c.#1#2#3{       {\it Riv. del Nuovo Cim. }{\bf #1} (19#2) #3}

\def\mpl#1#2#3{          {\it Mod. Phys. Lett. }{\bf #1} (19#2) #3}

\def\pc{private communication}

\relax
\documentstyle[12pt]{article}
\def\lsim{\:\raisebox{-0.5ex}{$\stackrel{\textstyle<}{\sim}$}\:}
\def\gsim{\:\raisebox{-0.5ex}{$\stackrel{\textstyle>}{\sim}$}\:}
\begin{document}
\begin{titlepage}
\begin{center}
\hfill{FTUV/94-36, IFIC/94-31}\\
\hfill TIFR/TH/94--25
\vskip 0.3cm
{\Large \bf LIMITS ON ASSOCIATED PRODUCTION OF VISIBLY
AND INVISIBLY DECAYING HIGGS BOSONS FROM Z DECAYS}\\
\vskip 1.0cm
{\large \bf F. DE CAMPOS}
\footnote{E-mail CAMPOSC at vm.ci.uv.es or 16444::CAMPOSC},
{\large \bf M. A. GARCIA-JARE\~NO}
\footnote{E-mail GARCIAMA at vm.ci.uv.es or 16444::GARCIAMA}\\
\vskip 0.5cm
{\large \bf ANJAN S. JOSHIPURA}
\footnote{Permanent address: Theory Group,
 Physical Research Lab., Ahmedabad, India},
{\large \bf J. ROSIEK}
\footnote{E-mail ROSIEK at vm.ci.uv.es or 16444::ROSIEK},
{\large \bf J. W. F. VALLE}
\footnote{E-mail VALLE at vm.ci.uv.es or 16444::VALLE}\\
\vskip .5cm
{\it Instituto de F\'{\i}sica Corpuscular - C.S.I.C.\\
Departament de F\'{\i}sica Te\`orica, Universitat de Val\`encia\\
46100 Burjassot, Val\`encia, SPAIN}\\
\vskip 0.5cm
{\large \bf D. P. ROY}
\footnote{E-mail dproy@theory.tifr.res.in }\\
\vskip .5cm
{\it Theoretical Physics Group,
Tata Institute of Fundamental Research, \\
Homi Bhabha Road, Bombay 400 005, INDIA}
\end{center}

\begin{abstract}
\baselineskip=11pt
{
Many extensions of the standard electroweak model
Higgs sector suggest that the main Higgs decay
channel is "invisible", for example, $h \to J J$
where $J$ denotes the majoron, a weakly interacting
pseudoscalar Goldstone boson associated to the
spontaneous violation of lepton number.
In many of these models the Higgs boson may
also be produced in association to a massive
pseudoscalar boson (HA), in addition to the standard
Bjorken mechanism (HZ).
We describe a general strategy to determine
limits from LEP data on the masses and couplings
of such Higgs bosons, using the existing data on
acoplanar dijet events as well as
data on four and six $b$ jet event topologies.
For the sake of illustration, we present constraints
that can be obtained for the ALEPH data.
}
\end{abstract}

\end{titlepage}

\section{Introduction}

The Higgs particle remains one of the missing links in the otherwise
well tested standard electroweak model (SM) \cite{HIGGS}. Its mass
is not fixed theoretically and experimental efforts to search for
the higgs boson, notably at LEP, have produced a lower bound
$\sim 63 $ GeV \cite{Aleph92} on its mass. This bound applies
to the standard model (SM) Higgs.

There exist well motivated extensions of the SM which are characterized
by a more complex Higgs structure than that of the SM \cite{hunter,cohen}.
These include ({\sl i}) the minimal supersymmetric standard model
(MSSM) ({\sl ii})
generic two or more Higgs doublet models and ({\sl iii}) the majoron type
models  characterized by a spontaneously broken global symmetry.
This symmetry could either be a lepton number \cite{CMP,fae},
 a combination of family lepton numbers or an R symmetry
in the context of supersymmetry \cite{MASI_pot3}. These extensions
contain one or more parameters in addition to the Higgs mass. Nevertheless,
a considerable region of the parameter space has already been ruled out
in these models using the LEP data \cite{aleph1}.

Among the three extensions mentioned above, the majoron type models are
qualitatively different compared to the other two as well as to the SM.
These models contain a massless goldstone boson, called majoron. There are
many types of majoron models \cite{fae}. Many of such models are
characterized by the spontaneous violation of a global $U(1)$
lepton number symmetry close to the electroweak scale.
This could also have important implications for the
structure of the electroweak phase transition and
the generation of the electroweak baryon asymmetry
\cite{ewbaryo}.

In such models the normal doublet
Higgs is expected to have sizeable invisible decay modes
to the majoron, due to the strong higgs
majoron coupling \cite{bj} - \cite{Joshi92a}.
This can have a significant effect on the Higgs phenomenology at LEP.
In particular, the invisible decay
could contribute to the standard signal looked for at LEP namely two
acoplanar jets and missing momentum. This feature of the majoron model
allows one to strongly constrain the Higgs mass \cite{alfonso,dproy,grivaz}
in spite of the occurrence of extra parameters compared to the standard model.
In particular, the limit on the predominantly doublet Higgs mass
can be seen to be close to the SM limit irrespective of the decay
mode of the Higgs boson \cite{alfonso,dproy,grivaz}.

Apart from giving  an interesting twist to the Higgs search strategy,
the  majoron models are quite well motivated theoretically \cite{fae}.
They allow the possibility of spontaneously generating lepton number violation
and hence neutrino masses. The Higgs couplings to majorons in these models
were discussed at length in \cite{JoshipuraValle92}.

As far as the neutral Higgs sector is concerned, all the models fall
 in two categories.  (A) Minimally extended SM characterized by the
 addition of a singlet Higgs: Such models are typical examples of
 mechanisms where the neutrinos obtain their mass at the tree level
 \cite{CON}.  (B) Models containing two Higgs doublets and a singlet
 under the $SU(2) \times U(1)$ group: These models though more complex
arise naturally in trying to generate the neutrino mass radiatively.
The Zee type \cite{zee} model modified to obtain spontaneous lepton
number violation is a typical example \cite{ewbaryo} in this category.
These models have the virtue of explaining smallness of neutrino masses
without invoking any high scale. Moreover, type B models are different
from the type A ones as far as the Higgs phenomenology is concerned
since they contain a massive pseudoscalar which can be produced in
association with the Higgs as in MSSM. All the existing analysis
of the invisible Higgs search \cite{alfonso,dproy,grivaz}
have concentrated on type A models and hence on the Bjorken process.

In this letter we extend this analysis to include the type (B)
models and particularly the associated production of the pseudoscalar
Higgs boson.
We describe a general strategy to determine
limits from LEP data on the H and A masses and
couplings using the existing data on acoplanar
dijet events. We also show how one can set
decay mode-independent limits on H and A masses
and couplings by including data on four and
six b jet events.
For the sake of illustration, we display the constraints
that can be obtained for the published ALEPH data.

\section{Higgs Boson Production and Decay}

Let us first recall the salient features of the type (B) models.
The Higgs sector is characterized by two doublets $\phi_{1,2}$ and a
singlet $\sigma$. The part of the scalar potential containing the
neutral Higgs fields is given in this case by
\begin{eqnarray}
\label{N2}
V &=&\mu_{i}^2\phi^{\dagger}_i\phi_i+\mu_{\sigma}^2\sigma^{\dagger}
\sigma
+ \lambda_{i}(\phi^{\dagger}_i\phi_i)^2+
   \lambda_{\sigma}
(\sigma^{\dagger}\sigma)^2+\nonumber\\
& &\lambda_{12}(\phi^{\dagger}_1\phi_1)(\phi^{\dagger}_2\phi_2)
+\lambda_{13}(\phi^{\dagger}_1\phi_1)(\sigma^{\dagger}\sigma)+
\lambda_{23}(\phi^{\dagger}_2\phi_2)(\sigma^{\dagger}\sigma)\nonumber\\
& & +\delta(\phi^{\dagger}_1\phi_2)(\phi^{\dagger}_2\phi_1)+
    \frac{1}{2}\kappa[(\phi^{\dagger}_1\phi_2)^2+\;\;h.\;c.]
\end{eqnarray}
where a sum over repeated indices $i$=1,2 is assumed.
Here $\phi_{1,2}$ are the doublet fields and $\sigma$
corresponds to the singlet carrying nonzero lepton number.

In writing down the above equation, we have imposed a discrete
symmetry $\phi_2\rightarrow-\phi_2$ needed to obtain natural
flavour conservation in the presence of more than one Higgs
doublet. For simplicity, we assume all the couplings and VEVS
to be real.

After minimization of the above potential
one can work out the mass matrix for the Higgs fields.
To this end we shift the fields as ($i$=1,2)
\begin{equation}
\label{shift1}
\phi_i=\frac{v_i}{\sqrt{2}}+\frac{R_i+iI_i}{\sqrt{2}} \nonumber
\end{equation}
\begin{equation}
\label{shift2}
\sigma=\frac{v_3}{\sqrt2}+\frac{R_3+iI_3}{\sqrt{2}}.
\end{equation}
The masses of the CP even fields $R_a$ ($a$=1...3) are obtained from
\begin{equation}
{\cal L}_{mass}=\frac{1}{2}R^T\;M_R^2\;R
\end{equation}
with
\begin{equation}
\label{MR}
M_R^2=\left (
\begin{array}{ccc}
2\lambda_1v_1^2&(\kappa+\lambda_{12}+\delta)v_1v_2&\lambda_{13}v_1v_3\\
(\kappa+\lambda_{12}+\delta)v_1v_2&2\lambda_2v_2^2&\lambda_{23}v_2v_3\\
\lambda_{13}v_1v_3&\lambda_{23}v_2v_3&2\lambda_3v_3^2\\
\end{array}  \right).
\end{equation}
The physical mass eigenstates $H_a$ are related to the corresponding
weak eigenstates as
\begin{equation}
\label{me}
H_a=O_{ab}\;R_b
\end{equation}
where,  $O$ is a 3$\times$3 matrix diagonalizing $M_R^2$
\begin{equation}
\label{diag}
O\;M_R^2\;O^T=diag \: (M_1^2,M_2^2,M_3^2).
\end{equation}
It is convenient to parametrize the matrix $O$ as
\begin{equation}
\label{O}
O=\left( \begin{array}{ccc}
     c_{\alpha} c_{\theta}&
     -s_{\alpha}c_{\omega}-c_{\alpha}s_{\omega}s_{\theta}&
     s_{\alpha}s_{\omega}-c_{\alpha}c_{\omega}s_{\theta}\\
s_{\alpha} c_{\theta}&c_{\alpha}c_{\omega}-s_{\alpha}s_{\omega}s_{\theta}&
    - c_{\alpha}s_{\omega}-s_{\alpha}c_{\omega}s_{\theta}\\
 s_{\theta} & s_{\omega} c_{\theta}&
     c_{\omega}c_{\theta}\\
\end{array}  \right)
\end{equation}
where $c_{\alpha} \equiv \cos \alpha $ etc.

In addition, there exists also a massive CP odd state A,
related to the doublet fields as follows
\begin{equation}
A=\frac{1}{V}(v_2I_1-v_1I_2).
\end{equation}
Its mass is given by
\begin{equation}
M_A^2=-\kappa V^2
\end{equation}
where $ V=(v_1^2+v_2^2)^{1/2}$.
Thus, after spontaneous $SU(2) \times U(1) \times U(1)_L$
breaking, we have a total of three massive CP even scalars
$H_{i} \:$ (i=1,2,3), plus a massive pseudoscalar $A$ and
the massless majoron $J$, simply given as $J=I_3$.

At an $e^+ e^-$ collider there are two main production mechanisms
that allow the production of the higgs boson through their
couplings to $Z$.
The relevant couplings for their production through
Bjorken process (HZ) are given as follows ($a$=1...3)
\begin{equation}
\label{HZZ2}
{\cal
L}_{HZZ}=(\sqrt{2}G_F)^{1/2}M_Z^2\;Z_{\mu}Z^{\mu}[\frac{v_1}{V}O_{1a}+
\frac{v_2}{V}O_{2a}]H_a.
\end{equation}
As long as the mixing appearing \eq{HZZ2} is $\O$(1), the Higgs
bosons can have significant couplings and hence appreciable
production rates through the Bjorken process.

In addition, the $H_a$ can also be produced in
association with the CP odd field $A$ through the HA coupling
\begin{equation}
\label{HAZ2}
{\cal L}_{HAZ}=-\frac{g}{cos\theta_W}Z^{\mu}\left [
\frac{v_2}{V}O_{1a}-\frac{v_1}{V}O_{2a}\right]H_a
\stackrel{\leftrightarrow}{\partial^\mu}\;A.
\end{equation}

In what follows we assume that at LEP only the
the lightest of the CP even Higgs boson, denoted
by $H_1 \equiv H$ can be accessible. Using the
matrix $O$ of \eq{O} in \eq{HZZ2} and \eq{HAZ2}
one gets the couplings of H to ZZ and HA as
\beq
\label{HZZ3}
{\cal
L}_{HZZ}=(\sqrt{2}G_F)^{1/2}M_Z^2\;Z_{\mu}Z^{\mu} \cos(\beta-\alpha)
\cos\theta H
\eeq
where $\tan \beta \equiv \frac{v_2}{v_1}$ and
\begin{equation}
\label{HAZ3}
{\cal L}_{HAZ}=-\frac{g}{cos\theta_W}Z^{\mu} \sin(\beta-\alpha)
\cos\theta\;H
\stackrel{\leftrightarrow}{\partial^\mu}\;A.
\end{equation}

As expected, one recovers the well known expressions
for the two doublets case in the limit $\theta\rightarrow 0$.
In particular, note that for sizeable value of $\cos \theta$,
one cannot simultaneously
suppress the production of Higgs through \eq{HZZ3} and \eq{HAZ3}
if it is allowed kinematically.

Now we turn to Higgs boson decay. For Higgs boson
masses accessible at LEP energies the main CP even
Higgs decay modes are into bb and JJ. The coupling
of the physical Higgses to $J$ follows from \eq{N2}.
One can express this coupling entirely in terms of the masses
$M_a^2$ and the mixing angles in the matrix $O$
\begin{eqnarray}
\label{J2}
{\cal L}_J&=&\frac{1}{2}J^2(2\lambda_3v_3R_3+\lambda_{13}v_1R_1+
           \lambda_{23}v_2v_3R_2)\\
         &=&\frac{J^2}{2v_3}(M_R^2)_{3a}R_a\\
       &=&\frac{1}{2}(\sqrt{2}G_F)^{1/2} \tan \gamma (O^T)_{3a}M_a^2H_a J^2
\end{eqnarray}
where tan$\gamma\equiv\frac{V}{v_3}$.
We have made use of \eq{me} and \eq{diag} in writing the last line.

Note from \eq{J2} that the CP even scalar Higgs bosons $H_i$
couple strongly to a pair of majorons leading to the invisible
decay signature. In contrast to $H_i$, the pseudoscalar $A$ does
 not decay into one or three majorons since the couplings
$AJ^3$ or $AHJ$, although possible in general, do not exist
at tree level in our simplest model described above.
Because of the form of the scalar potential given in
\eq{N2} only even powers of the majoron field
$J \equiv R_3$ appear once the expansions
in \eq{shift1} and \eq{shift2} are used.
As a result the couplings $AJ^3$ and $AHJ$
are absent from the potential and therefore the
$A$ decays visibly to fermion antifermion pair
\footnote{Note that $A$ could decay invisibly at the
tree level through the process $A\rightarrow Z^* H\rightarrow
 \nu \overline{\nu} JJ$. However the branching ratio in this mode
is quite small.}.
The branching fraction $B_A$ for $A\rightarrow b \overline{b}$ is nearly one.
Deviation from unity of this branching ratio is model dependent.
If all fermions obtain their mass by coupling to
only one higgs boson, as in the models considered in
ref. \cite{Joshi92,ewbaryo}, then one has
\begin{equation}
B_A=\frac{1}{1+r}
\end{equation}
where
\begin{equation}
r \approx \sum_{f} \frac{m_f^2 (1- 4 m_f^2/m_A^2)^{1/2}}
{m_b^2 (1 - 4 m_b^2/m_A^2)^{1/2}}
\end{equation}
and the sum is over all the fermions except $b$.
The rate for $H\rightarrow b \overline b$ can be expressed as
\begin{equation}
\Gamma(H\rightarrow b \overline b)=\frac{3\sqrt 2 G_F}{8
\pi}M_Hm_b^2(1-4m_b^2/M_H^2)^{3/2} \cos^2 \alpha \cos^2 \theta
\end{equation}

On the other hand the  width for the invisible $H$ decay  can be
parametrized by
\begin{equation}
\label{HJJ}
\Gamma(H\rightarrow JJ)=\frac{\sqrt 2 G_F}{32 \pi} M_H^3 \:
(\tan \gamma\;\; O_{13})^2
\end{equation}
For many choices of parameters the invisible decay mode can be
rather important and, in fact, provide the strongest limits.
Finally, note that additional decay modes of $H$ to AA may also exist.

\section{Search Strategy and Analysis}

Since $A$ can decay only visibly, one expects dijets +
missing momentum as a signal of the Higgs production in our
simplest model. This signal arises from three processes -- i.e.
$Z^* \rightarrow  \nu \bar{\nu}$ with
$H \rightarrow b \overline{b}$,
or $Z^* \rightarrow q \overline{q}$ with $H \ra JJ$
or $H\rightarrow JJ$ with $A\rightarrow b \overline{b}$.
For each process one has a sizeable missing momentum which is aligned
neither along the beam nor along the jets.  In contrast, for the SM
background, the missing momentum arises from i) jet fluctuation
(including decay $\nu$ from $b,c$ and $\tau$ jets) in which case it is
aligned along the jets, and ii) initial state radiation (ISR) or the
two-photon process $e^+e^- \rightarrow (e^+e^-) \gamma\gamma$ in
which case it is aligned along the beam direction.  This enables one
to eliminate the SM background in the dijet + missing momentum channel
by a suitable combination of kinematic cuts without depleting the
signals seriously.  In fact, this procedure has been extensively used
to search for the SM Higgs signal (first process) in the LEP data and
obtain the corresponding mass limit \cite{Aleph92}.  More recently the
analysis has been extended to the invisibly decaying Higgs signal
(second process) and obtain the corresponding mass limits for the A
type majoron models \cite{alfonso,dproy,grivaz}.  In this section we
shall extend the analysis further to include the Higgs signal from the
third process of associated production (HA) and study the resulting mass
limits for the B type models.

We shall use a parton level Monte Carlo event generator, which has
been shown \cite{dproy} to reproduce the signals obtained with the
full Monte Carlo program \cite{aleph1} quite well.

For our illustrative purposes, below we will determine
the corresponding Higgs boson mass limits which can
be obtained from the ALEPH data sample of ref. \cite{Aleph92}
based on a statistics of $\sim  1.23$ million hadronic Z events.
A detailed account of the experimental cuts can
be found in \cite{aleph1}.  We shall only summarize the main features.
One starts with a visible mass cut, $M_{jj} < 70~{\rm GeV}$, to ensure a
sizeable missing energy ($E\!\!\!/$) and momentum ($\vec p\!\!\!/$).
A low angle cut, requiring the energy deposit within $12^\circ$ of the
beam axis to be $< 3$ GeV and that beyond $30^\circ$ to be $> 60\%$ of
the visible energy, removes jets close to be beam pipe where
measurement errors can simulate a large $E\!\!\!/$.  An acollinearity
cut, requiring the angle between the two jets to be $< 165^\circ$,
suppresses the $Z \rightarrow q\bar q$ and $\tau^+\tau^-$ background
where the $E\!\!\!/$ can come from jet fluctuation (including escaping
$\nu$).  Moreover an isolation cut on $\vec p\!\!\!/$ removes the
$E\!\!\!/$ background from the fluctuation of any one of the jets.  A
cut on the angle ($\alpha$) of $\vec p\!\!\!/$ with respect to the beam
axis, $\tan \alpha > 0.4$, suppresses the background from ISR and two
photon processes.  The $E\!\!\!/$ background coming from the jet
fluctuations along with an ISR are removed by an acoplanarity cut,
requiring the azimuthal angle between the two jets to be $<
175^\circ$.  An acoplanarity cut for 3-jet like events, requiring the
sum of the 3 dijet angles to be $< 350^\circ$, removes the $E\!\!\!/$
background arising from the fluctuation of these jets.  The remaining
few events are residual two-photon events, which are removed by a
total $p_T$ cut.  The cuts remove all the events from the data sample
\cite{aleph1}, while retaining $\gsim 50\%$ of the signal for each of
the three processes mentioned above.

We denote the number of signal events for the three processes, after
the cuts, by $N_{SM}, N_{JJ}$ and $N_A$ respectively, assuming no
suppression due to the mixing angles or branching fractions in each
case.  Then the expected number of signal events, after incorporating
these effects, is given by
\begin{equation}
N_{expt}=\epsilon_B^2\left[ B N_{J}+(1-B) N_{SM}\right]+\epsilon_A^2
B_A B N_A
\end{equation}
where
\begin{equation}
B = BR (H \ra JJ)
\end{equation}
is the branching fraction for the $H$ decaying into the invisible mode
and $\epsilon_A \equiv \sin (\beta - \alpha) \cos \theta$ and
$\epsilon_B \equiv \cos (\beta - \alpha) \cos \theta$.

The number of signal events for the three processes that pass the cuts
are shown in Fig. 1. As we can see $N_A \gg N_{SM}, \: N_{JJ}$ thus
implying that, if kinematically open and not suppressed by mixing
angles, associated production tends to give the strongest limits.  We
shall now consider the limits which can be obtained by comparing eq.
(22) with the 95\% CL limit of 3 events, corresponding to 0 events in
the data sample \cite{aleph1} after the cuts.

First, we assume that B=0. In this case we get the
limits corresponding to the SM decay mode. Note that this
is the weakest possible limit for the Bjorken
coupling strength $\epsilon_B^2$.
This is in complete agreement with the results of ref.
\cite{alfonso,dproy} as seen in Fig. 2.
Note that in this case we can set no limits on
$\epsilon_A^2$.

We can also obtain limits assuming B=1. From Fig. 1
one can see that, except for the region very close
to the edge of the phase space, $m_A + m_H \sim m_Z$,
the major contribution comes from the associated production
($N_A$) allowing the corresponding coupling strength parameter
$\epsilon_A^2$ to be strongly constrained, as seen
in Fig. 3. It is easily seen that the limits close to the
edge of the phase space are typically at the level
$\epsilon_A^2 \sim 0.05$. However, very close to
the kinematical limit the Bjorken process becomes
important and could further strengthen the limits
we have obtained.
Outside the region where associated production
is kinematically possible, the Bjorken process can
still take place and the limits on $\epsilon_B^2$ are
the same as obtained in ref. \cite{alfonso,dproy,grivaz}.
The limits on the coupling strengths $\epsilon^2_A$ and
$\epsilon^2_B$ can be improved further by using more
recent data from LEP.

In the case B=0 considered earlier one does not get any limit on
$\epsilon_A^2$ since in this case the associated production does not
lead to the dijets + missing momentum signal, we have considered so
far. However one could derive limits on the $\epsilon_A^2$ using data
on four and six b jets topologies, assuming good b identification.
As an illustration of what can be achieved we assume some typical
value for the corresponding branching ratios
$BR( e^+ e^- \rightarrow H\:A \rightarrow 4b) \leq \: 5 \times 10^{-4}$.
Using this value we can determine the excluded region for different values
of $\epsilon^2_{A}$ as a function of $m_A$ and $m_H$ from
\beq
\label{rich}
BR_{HA}=\frac{1}{2}\:(BR)_{\nu \nu} \lambda^3\:\epsilon^2_{A} \:,
\eeq
where ${BR}_{\nu \nu}$ is the Z decay branching ratio into
one generation of neutrinos in the SM, the factor 1/2 refers to the
scalar HA decay mode and $\lambda$ is the corresponding phase space
factor
$$
\lambda(s,m_H,m_{A})=
\frac{\sqrt{(s+m_{H}^{2}-m_{A}^{2})^2-4\:s\:m_{H}^2}}{s}
$$
The resulting excluded regions are shown in Fig. 4.
On the other hand if H has a mass bigger than
$2m_A$ then \eq{rich} would give a weaker limit
since it should now be multiplied by the branching ratio
for H into 2b. But now additional information can be obtained
from the study of events with six b jet topologies.
One may obtain a similar limit using an illustrative
reference limit for the corresponding branching
$BR( e^+ e^- \rightarrow H\:A \rightarrow 6b) \leq 4 \times \:10^{-4}$.

\section{Discussion}

Here we have shown that the production of invisibly decaying
Higgs bosons in Z decays through the associated channel HA
can lead to substantial limits on Higgs boson masses and couplings.
Such invisibly decaying Higgs bosons arise in
a wide class of \21 majoron-type models, such as the majoron
extension of the Zee model for radiatively induced \neu masses,
as well as extensions of the minimal supersymmetric standard
model with spontaneously broken R parity \cite{MASI_pot3}.

Note that it is also possible to derive a truly
decay mode independent limit on the H mass in the
case of associated HA production by allowing B to vary
from 0 to 1 and by combining
the data on dijet + missing momentum, with those from four
and six b jet searches described above.

It is important to update the limits obtained here by
using the full statistics provided by the four LEP
experiments and take them into account in designing the
strategies to search for the Higgs boson at higher
energies, such as at the LHC \cite{kane} and NLC
\cite{EE500}.

Finally we note that a similar analysis may be performed
in the context of the minimal supersymmetric standard model,
where the invisible decay of the Higgs bosons into neutralinos
may take place.
A major difference in the latter case arises from the fact
that in the MSSM case the pseudoscalar boson A is the one
most likely \cite{Zerwas} to decay invisibly, whereas in the
case discussed in the present paper the pseudoscalar
A always decays into $b\bar{b}$.

{\bf Acknowledgements}

We thank Francois Richard for useful discussions
during a visit to LAL at Orsay (F. de C. and J.W.F.V.).
J.W.F.V. also wishes to express gratitude to the
organizers of WHEPP-3 workshop held in Madras
for the hospitality extended to him in January 94.
This visit initiated the collaboration with D.P.R.
This work was supported by DGICYT under grant number
PB92-0084 and SAB94-0014 (A. S. J.), as well as a postdoctoral
fellowship (J.R.) and an FPI fellowship (M.A.G.J.). The work
of F. de Campos was supported by CNPq (Brazil).

\newpage

\noi
{\bf Figure Captions}\\
{\bf Figure 1}\\
Numbers of expected
dijet+missing momentum events for the processes $e^+ e^- \rightarrow
Z^*H \rightarrow \nu \bar{\nu}$  $b\bar b$ (N$_{SM}$), $e^+ e^-
\rightarrow Z^*H \rightarrow q \bar q $ JJ (N$_{JJ}$) and $e^+ e^-
\rightarrow H\:A \rightarrow J\: J b\bar{b}$ (N$_{HA}$) after imposing
ALEPH cuts \cite{aleph1,Aleph92}.\\
{\bf Figure 2}\\
Illustrative limits on $\epsilon^2_{B}$ for visibly decaying
H (B=0) from the $ Z^* H \rightarrow \nu \nu$ + two jets channel
as a function of $m_H$ using the ALEPH data of ref. \cite{Aleph92}.\\
{\bf Figure 3}\\
Limits on $\epsilon^2_{A}$ in the $m_A m_H$ plane, based on the
 $e^+ e^- \rightarrow H \:A \rightarrow J\:J b\bar{b}$ production channel.
We have assumed BR ($H \rightarrow J\:J$) = 100\% and a visibly
decaying A with branching ratio into bb given in eq. (18)\\
{\bf Figure 4}\\
Limits on $\epsilon^2_{A}$ for the B=0 case based only on
 $e^+ e^- \rightarrow H\:A \ra b \bar{b} b \bar{b}$ channel.
We have assumed a hypothetical sensitivity for the $4b$ channel
of BR($e^+ e^- \rightarrow H\:A \rightarrow 4b) \leq 5\times\:10^{-4}$.
\\

\end{document}